%% file: main.tex
\documentclass[11pt]{article}

\usepackage[font=libertinus,citestyle=numeric,status={}]{kurbanlab}
\input{affiliations}
\graphicspath{{figures/}}

\newcommand{\tabsizeM}{\fontsize{10bp}{12bp}\selectfont}

\newcommand{\notesize}{\fontsize{9bp}{11bp}\selectfont}

\newcommand{\thead}[1]{\multicolumn{1}{c}{#1}}

\newcommand{\snote}[1]{Supplementary Note ``\nameref{#1}''}

\newcounter{suppnote}
\makeatletter
\newcommand{\suppnote}[1]{\refstepcounter{suppnote}\section*{Supplementary Note: #1}\addcontentsline{toc}{section}{Supplementary Note: #1}\def\@currentlabelname{#1}}
\makeatother
\newcommand{\beginsupplement}{\setcounter{figure}{0}\renewcommand{\thefigure}{S\arabic{figure}}\renewcommand{\theHfigure}{S\arabic{figure}}\setcounter{table}{0}\renewcommand{\thetable}{S\arabic{table}}\renewcommand{\theHtable}{S\arabic{table}}\renewcommand{\figurename}{Supplementary Fig.}\renewcommand{\tablename}{Supplementary Table}}
\title{The parity gap in crystal tensor prediction}
\RunningTitle{The parity gap in crystal tensor prediction}
\Author[orcid=0000-0002-1458-302X]{Can Polat}{tamu}
\Author[corresponding=kurbanm@ankara.edu.tr,orcid=0000-0002-7263-0234]{Mustafa Kurban}{tamuq,ankara}
\Author[orcid=0000-0001-9069-770X]{Erchin Serpedin}{tamu}
\Author[corresponding=hkurban@hbku.edu.qa,orcid=0000-0003-3142-2866]{Hasan Kurban}{hbku}
\CodeURL{https://github.com/KurbanIntelligenceLab/equiparity}
\Venue{Preprint}

\begin{document}
\maketitle

\begin{abstract}
Crystal symmetry dictates whether a physical response tensor must vanish, establishing a direct test for machine learning predictions independent of property calculations. We derive the \emph{parity gap}, a group-theoretic metric quantifying the piezoelectric tensor freedom permitted by a crystal's proper rotation subgroup $SO(3)$ but eliminated by inversion symmetry in $O(3)$. Across state-of-the-art equivariant neural network architectures, unconstrained $SO(3)$ models systematically predict forbidden non-zero responses matching the parity gap of each centrosymmetric crystal class, while polar distortion paths dynamically map output responses to the loss of inversion symmetry. Regression controls confirm that enforcing full $O(3)$ parity incurs no consistent accuracy cost across predictive tasks. Crucially, while training interventions using explicit zero labels reduce violation magnitudes, they leave residual forbidden outputs. Exact physical compliance instead requires structural enforcement through $O(3)$ representation design or explicit output antisymmetrization. The parity gap thus provides a unified framework to distinguish empirical error reduction from exact structural compliance with physical law.
\end{abstract}

\printkeywords

\input{sections/introduction}
\input{sections/results}
\input{sections/discussion}
\input{sections/methods}

\begin{availability}
All datasets supporting this study are public. Molecular targets are from QM9~\cite{ramakrishnan2014qm9}; crystal structures are from the Materials Project~\cite{jain2013materials}; piezoelectric and elasticity tensors are from the corresponding Materials Project repositories~\cite{dejong2015piezo,dejong2015elastic}. Repository manifests provide content hashes, split definitions, benchmark identifiers, and machine-readable records underlying the figures and tables. Project data and software are archived at \href{https://doi.org/10.5281/zenodo.22003285}{doi:10.5281/zenodo.22003285}. Experimental software is available at \url{https://github.com/KurbanIntelligenceLab/equiparity} under the MIT license.
\end{availability}

\begin{funding}
This work received no specific funding from any funding agency in the public, commercial, or not-for-profit sectors.
\end{funding}

\begin{contributions}
C.P. conceived the study, designed and implemented the matched-pair parity-ablation and verification pipeline, conducted all computational experiments and benchmark runs, performed the data analysis, and wrote the original manuscript. M.K., E.S., and H.K. supervised the research, contributed to the interpretation of results, and reviewed and edited the manuscript.
\end{contributions}

\begin{conflicts}
The authors declare no competing financial or non-financial interests.
\end{conflicts}

\bibliography{references}

\clearpage
\appendix
\section{Supplementary Information}
\beginsupplement
\input{sections/supplementary}
\FloatBarrier

\end{document}

%% file: affiliations.tex
\DeclareAffiliation{hbku}{%
  College of Science and Engineering, Hamad Bin Khalifa University, Doha, Qatar}

\DeclareAffiliation{tamu}{%
  Department of Electrical and Computer Engineering,
  Texas A\&M University, College Station, Texas, USA}

\DeclareAffiliation{tamuq}{%
  Department of Electrical and Computer Engineering,
  Texas A\&M University at Qatar, Doha, Qatar}

\DeclareAffiliation{ankara}{%
  Department of Prosthetics and Orthotics,
  Ankara University, Ankara, Turkey}

\DeclareAffiliation{iub}{%
  Luddy School of Informatics, Computing, and Engineering,
  Indiana University Bloomington, Bloomington, IN, USA}


%% file: sections/introduction.tex
\section*{Introduction}

Neumann's principle dictates that a crystal property tensor must remain invariant under the material's point group; consequently, spatial inversion forces all odd-rank polar tensors—such as piezoelectricity—to vanish identically on centrosymmetric structures, independent of chemical composition~\cite{dejong2015piezo,nye1985crystals}. Verifying these symmetry-forced zeros provides a crucial benchmark for the physical validity of machine learning models alongside standard regression accuracy~\cite{yan2024gmtnet}. Equivariant architectures now target diverse crystal response properties: MatTen predicts elastic tensors~\cite{wen2024matten}, GMTNet models dielectric, piezoelectric, and elastic tensors~\cite{yan2024gmtnet}, and EATGNN captures complete piezoelectric tensors using equivariant attention~\cite{dong2025eatgnn}. However, geometric frameworks like Tensor Field Networks often enforce only proper-rotation symmetry ($\mathrm{SO}(3)$), leaving the parity constraints imposed by spatial inversion unenforced~\cite{thomas2018tfn,duval2024hitchhiker}.

The key mathematical distinction lies between the proper-rotation group $\mathrm{SO}(3)$ and the full orthogonal group $\mathrm{O}(3)$, which incorporates improper rotations and inversion~\cite{brandstetter2021geometric}. In an $\mathrm{O}(3)$-equivariant representation, features explicitly specify both angular degree and inversion parity, directly constraining valid tensor-product couplings~\cite{geiger2022e3nn,polat2026cliffordip}. While modern architectures such as NequIP~\cite{batzner2022nequip}, Allegro~\cite{musaelian2023allegro}, and MACE~\cite{batatia2022mace} support full $\mathrm{O}(3)$ enforcement, improper-rotation constraints are not universally applied. Proper rotations alone can sometimes forbid response tensors. For example, cubic class 432 eliminates piezoelectricity purely through rotational symmetry~\cite{klapper2006point}. For any equivariant mapping, inputs fixed under a point group must map to outputs residing within the corresponding invariant subspace~\cite{kaba2023symmetry}, though post-hoc group averaging offers an alternative route to enforce output transformation laws over flexible internal features~\cite{puny2022frame}.

Despite these theoretical principles, the extent to which $\mathrm{SO}(3)$-equivariant models preserve residual non-physical freedom in centrosymmetric crystals remains unquantified~\cite{polat2026cliffordstf}. We address this gap by formulating the \emph{parity gap}, a metric that counts unconstrained tensor degrees of freedom across crystal classes while maintaining physical index symmetries. Through prediction experiments across core neural architectures, we isolate the operational impact of parity enforcement, evaluating whether unconstrained models generate non-physical outputs and distinguishing explicit structural constraints from soft loss-based training interventions. Our contributions are:
\begin{itemize}
\item \textbf{Deriving the parity-gap criterion:} Quantifies the unconstrained inversion freedom for polar tensors across centrosymmetric crystal classes, identifying where proper rotations fully determine symmetry constraints.
\item \textbf{Conducting matched architectural ablations:} Isolates parity manipulations across three neural-network backbones to evaluate symmetry-forced zeros alongside molecular and crystal regression controls while tracking architectural changes.
\item \textbf{Distinguishing empirical learning from exact physical compliance:} Demonstrates through point-group analysis, continuous polar-distortion paths, and output antisymmetrization that soft zero-label supervision reduces violation magnitudes but fails to eliminate residual non-physical responses.
\end{itemize}

%% file: sections/results.tex
\section*{Results}

\subsection*{A group-theoretic criterion for forbidden responses}

The parity gap measures the residual tensor freedom permitted under proper rotation symmetry $\mathrm{SO}(3)$ before enforcing spatial inversion in $\mathrm{O}(3)$. A vanishing gap indicates that proper rotations alone fully constrain the tensor space to zero on centrosymmetric inputs, whereas a positive gap quantifies the exact unconstrained subspace eliminated by inversion.

Let $f_\theta(x)$ represent the predicted tensor for a periodic crystal structure $x$ parameterized by $\theta$. The target space $V$ carries the polar-tensor representation $D(Q)=Q^{\otimes r}$ restricted to its physical index symmetries~\cite{zaverkin2024ictp}. For piezoelectricity, setting $r=3$ yields an eighteen-dimensional space $V=\mathbb{R}^{3}\otimes\mathrm{Sym}^{2}(\mathbb{R}^{3})$~\cite{dong2025eatgnn}, where the Frobenius norm $\lVert f_\theta(x)\rVert_F$ measures the output magnitude in physical units. Model outputs are analyzed under three baseline invariances: 
\begin{itemize}
    \item \textbf{(A1)} species-preserving atomic permutation invariance.
    \item \textbf{(A2)} invariance under rigid translations and individual lattice-vector shifts.
    \item \textbf{(A3)} transformation equivariance $f_\theta(Q\!\cdot\!x)=D(Q)f_\theta(x)$ for an enforced symmetry group $\mathcal{G}$,
\end{itemize}
where we denote $G$ as the crystal point group and $G^{+}=G\cap\mathrm{SO}(3)$ as its proper rotation subgroup. The translation condition (A2) encompasses fractional shifts associated with non-symmorphic space-group operations.

On any centrosymmetric crystal, spatial inversion yields a periodic input identical up to transformations covered by (A1)--(A2), forcing the structural identity $f_\theta(I\!\cdot\!x)=f_\theta(x)$. For odd-rank polar tensors, full $\mathrm{O}(3)$ equivariance simultaneously requires sign reversal under inversion ($f_\theta(I\!\cdot\!x)=-f_\theta(x)$), jointly forcing the output to vanish. Conversely, enforcing only proper rotations leaves the prediction unconstrained within the invariant subspace $V^{G^{+}}$.

\begin{theorem}[Parity gap]\label{thm:gap}
Let $x$ be centrosymmetric with point group $G$, and let $V$ be a polar rank-$r$ tensor space restricted to its physical index symmetries. Under (A1)--(A2), O(3) equivariance forces $f_\theta(x)=0$ for odd $r$ at every $\theta$, whereas SO(3) equivariance forces membership in $V^{G^{+}}$. The parity gap
\[
\gamma_r(G)=\dim V^{G^{+}}-\dim V^G
\]
is zero for even $r$ and equals $\dim V^{G^{+}}$ for odd $r$.
\end{theorem}

The coset decomposition $G=G^{+}\sqcup I G^{+}$ establishes that the parity gap depends strictly on group action rather than model parameterization (full proof detailed in \snote{snote:proofs}). Spatial inversion imposes no additional constraints on even-rank polar tensors, but completely eliminates the proper-rotation invariant subspace for odd-rank tensors. 

As shown in Table~\ref{tab:gap}, ten of the eleven centrosymmetric crystal classes exhibit a positive piezoelectric parity gap. The gap vanishes exclusively for class $\mathrm{m}\bar{3}\mathrm{m}$, whose proper subgroup 432 forbids piezoelectricity purely through rotational symmetry. Restricting calculations to physical index symmetries excludes the fully antisymmetric rank-3 tensor otherwise preserved by proper rotations~\cite{nye1985crystals}. Consequently, positive gap values permit non-zero outputs within $V^{G^{+}}$, establishing an empirical test for whether trained neural networks exploit this residual freedom.

\begin{table}[!hbp]
\centering
\caption{\textbf{Parity gap values across centrosymmetric crystal classes.} The parity gap counts the additional inversion constraints governing centrosymmetric polar tensors. The rank-3 column evaluates the eighteen-dimensional piezoelectric tensor space (symmetric in its strain indices), while the rank-1 column corresponds to the polar-vector space. $G^{+}$ denotes the proper-rotation subgroup. A vanishing parity gap ($\gamma_r = 0$) indicates that proper rotations alone fully force a zero response on that tensor space.}
\label{tab:gap}
\tabsizeM
\begin{tabular*}{\textwidth}{@{}l l l@{\extracolsep{\fill}}
  S[table-format=2.0] S[table-format=2.0]@{}}
\toprule
& & & \multicolumn{2}{c}{Parity gap $\gamma_r$ at odd rank} \\
\cmidrule(l){4-5}
Class $G$ & $G^{+}$ & Crystal system
 & \thead{$r=1$} & \thead{$r=3$} \\
& & & \thead{polarization} & \thead{piezoelectric} \\
\midrule
$\bar{1}$        & $1$   & triclinic     &  3 & 18 \\
$2/\mathrm{m}$            & $2$   & monoclinic    &  1 &  8 \\
$\mathrm{mmm}$            & $222$ & orthorhombic  &  0 &  3 \\
$4/\mathrm{m}$            & $4$   & tetragonal    &  1 &  4 \\
$4/\mathrm{mmm}$          & $422$ & tetragonal    &  0 &  1 \\
$\bar{3}$        & $3$   & trigonal      &  1 &  6 \\
$\bar{3}\mathrm{m}$       & $32$  & trigonal      &  0 &  2 \\
$6/\mathrm{m}$            & $6$   & hexagonal     &  1 &  4 \\
$6/\mathrm{mmm}$          & $622$ & hexagonal     &  0 &  1 \\
$\mathrm{m}\bar{3}$       & $23$  & cubic         &  0 &  1 \\
$\mathrm{m}\bar{3}\mathrm{m}$      & $432$ & cubic         &  0 &  0 \\
\bottomrule
\end{tabular*}
\end{table}

\subsection*{Matched models differ on symmetry-forced zeros}

To isolate parity enforcement across architectures, we constructed matched $\mathrm{O}(3)$ and $\mathrm{SO}(3)$ variants of NequIP, Allegro, and MACE using unified datasets, splits, hyperparameters, and random seeds. The $\mathrm{O}(3)$ models assign natural spherical-harmonic parity, whereas their $\mathrm{SO}(3)$ counterparts relabel edge harmonics and hidden features as parity-even. Relabeling expands permitted tensor-product paths and alters parameter counts while preserving angular degrees. Full architectural specifications, numerical verification probes, and regression baselines are detailed in \snote{snote:construction} and Supplementary Table~\ref{stab:accuracy}.

Model performance was evaluated on 2,000 centrosymmetric insulating crystals selected from the Materials Project~\cite{jain2013materials} and verified via spglib~\cite{togo2024spglib}, completely disjoint from training and validation sets. Evaluation targeted both idealized coordinates and density-functional-relaxed geometries, tracking predicted tensor norm departures from zero. Non-physical predictions are quantified by the false-flag fraction exceeding an operating threshold of $\tau=0.01$\,C\,m$^{-2}$.

\begin{figure}[!t]
\centering
\includegraphics[width=0.68\textwidth]{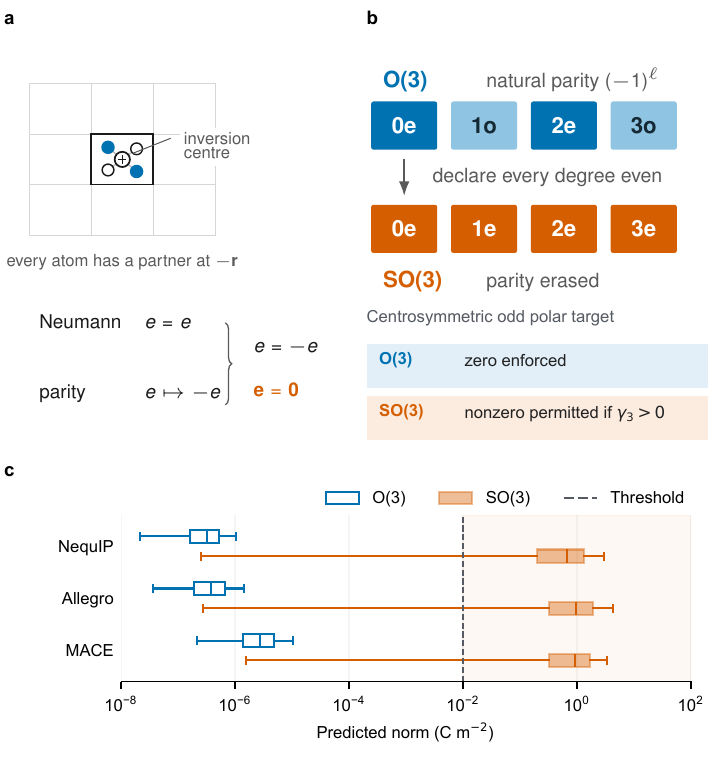}
\caption{\textbf{Parity enforcement separates matched models on symmetry-forced zeros.} \textbf{a}, Spatial inversion maps a centrosymmetric crystal to an equivalent periodic representation, requiring odd-rank polar tensors to be simultaneously invariant and sign-reversed, forcing a physical zero. \textbf{b}, Equivariant $\mathrm{O}(3)$ architectures assign natural spherical-harmonic parity $(-1)^l$, whereas $\mathrm{SO}(3)$ partners relabel edge harmonics and hidden features as parity-even ($e$), opening additional tensor-product pathways and permitting non-zero responses when the parity gap is positive. \textbf{c}, Predicted piezoelectric norms on 2,000 idealized centrosymmetric test crystals across three matched architectural pairs (averaged over three seeds). Boxes span 25th to 75th percentiles; central lines represent medians; whiskers denote 5th to 95th percentiles. Open blue boxes denote $\mathrm{O}(3)$ models; filled orange boxes denote $\mathrm{SO}(3)$ models. The dashed vertical line marks the operating false-flag threshold ($\tau=0.01$\,C\,m$^{-2}$). The low $\mathrm{SO}(3)$ tail comprises zero-gap crystals where proper rotations alone mandate zero.}
\label{fig:headline}
\end{figure}

Parity enforcement creates a sharp divergence on symmetry-forced zeros across all tested architectures (Fig.~\ref{fig:headline}). Every matched $\mathrm{O}(3)$ model maintains predicted norms below the operating threshold across all seeds on idealized structures, returning medians consistent with numerical precision limits. In contrast, unconstrained $\mathrm{SO}(3)$ models produce high false-flag fractions averaging 89.533\% (NequIP), 90.950\% (Allegro), and 90.767\% (MACE), generating forbidden outputs with magnitudes comparable to non-centrosymmetric physical tensors in the training set.

The high violation rates of $\mathrm{SO}(3)$ models remain stable across relaxed and idealized geometries. Conversely, $\mathrm{O}(3)$ compliance strictly mirrors the underlying coordinate symmetry: one raw structure exceeding threshold in three $\mathrm{O}(3)$ runs refined to a non-centrosymmetric space group under tighter tolerances, while an idealized structure failing tight checks remained below threshold across all $\mathrm{O}(3)$ arms. These edge cases delineate theoretical symmetry guarantees from numerical coordinate tolerances (further threshold sweeps and statistical summaries are compiled in \snote{snote:robustness}, Supplementary Fig.~\ref{sfig:thresholds}, and Supplementary Table~\ref{stab:pooling}).

Paired experiments comparing extensive summed readouts to mean-pooled configurations confirm that pooling choice does not drive this separation. Recalibrating thresholds for mean-pooled outputs preserves zero false-flag fractions in $\mathrm{O}(3)$ models and persistent violations in $\mathrm{SO}(3)$ arms. Furthermore, evaluating an unpaired EquiformerV2 (EqV2) implementation~\cite{liao2024equiformerv2} yields a 95.700\% false-flag rate; detected rotation and determinism defects place this core outside exact $\mathrm{SO}(3)$ assumptions.

\subsection*{The discrepancy follows crystal symmetry}

\begin{figure}[!t]
\centering
\includegraphics[width=0.66\textwidth]{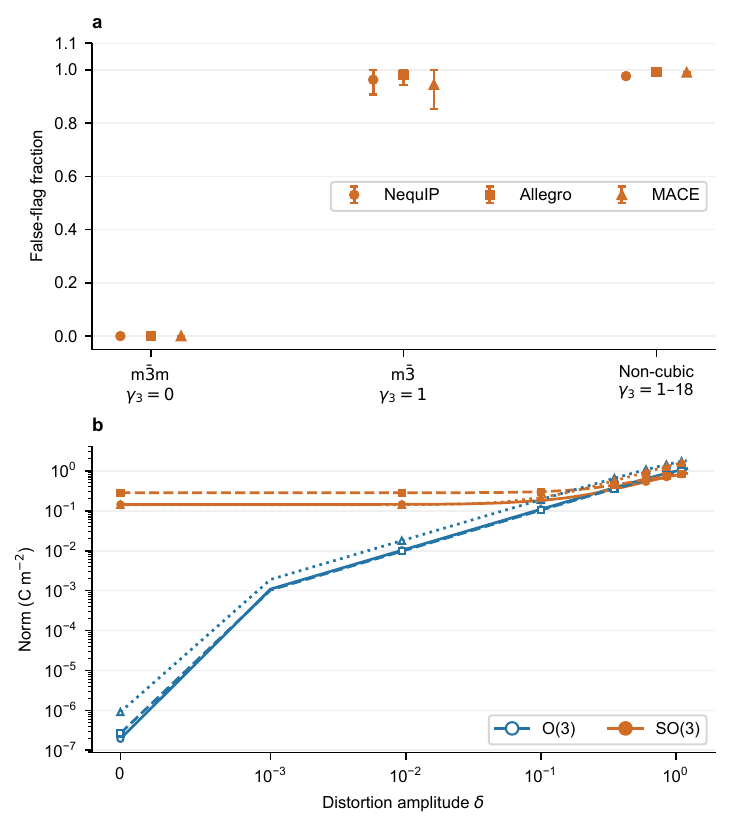}
\caption{\textbf{Prediction patterns follow structural crystal symmetries.} \textbf{a}, False-flag fractions for matched $\mathrm{SO}(3)$ models broken down by point-group family. The zero-gap $\mathrm{m}\bar{3}\mathrm{m}$ family mandates zero under proper rotations alone, whereas positive-gap families permit non-zero outputs within allowed invariant subspaces. Points indicate three-seed means; error bars display 95\% percentile-bootstrap confidence intervals across structures. All matched $\mathrm{O}(3)$ models yield zero false-flag fractions across all families. \textbf{b}, Predicted piezoelectric norm along a continuous polar-distortion path in rutile TiO$_2$ (averaged over three seeds). Open blue markers denote $\mathrm{O}(3)$ models; filled orange markers denote $\mathrm{SO}(3)$ models (circles, squares, and triangles denote NequIP, Allegro, and MACE, respectively). Non-zero distortion amplitudes explicitly break spatial inversion symmetry, allowing $\mathrm{O}(3)$ predictions to scale continuously above the arithmetic floor.}
\label{fig:mechanism}
\end{figure}

The distribution of $\mathrm{SO}(3)$ output violations across space groups directly reflects the parity-gap criterion. On the 166 evaluation crystals belonging to class $\mathrm{m}\bar{3}\mathrm{m}$, where proper rotations alone mandate a zero tensor, all matched $\mathrm{SO}(3)$ architectures remain below the operating threshold (Fig.~\ref{fig:mechanism}a). Conversely, $\mathrm{SO}(3)$ models systematically generate non-zero outputs across positive-gap classes where proper subgroups permit tensor freedom. The subset of positive-gap structures remaining below threshold reflects the freedom of model parameters within the allowed invariant subspace.

Point-group family breakdown further validates this structural dependence. The 18 crystals in cubic class $\mathrm{m}\bar{3}$ (carrying a 1D gap) and various non-cubic classes (gaps up to 18D) exhibit high violation rates, contrasting sharply with zero-gap $\mathrm{m}\bar{3}\mathrm{m}$ structures. Thus, proper subgroup symmetries govern the presence of forbidden predictions.

Continuous lattice distortions confirm this physical transition. Displacing atoms in centrosymmetric rutile TiO$_2$ (space group $4/\mathrm{mmm}$, positive parity gap) along a polar mode across 33 verified amplitudes traces model responses across the symmetry boundary (Fig.~\ref{fig:mechanism}b). At zero distortion, $\mathrm{O}(3)$ models predict near-zero norms while $\mathrm{SO}(3)$ models output non-zero tensors. As polar displacement breaks inversion symmetry, $\mathrm{O}(3)$ predictions continuously scale up from the arithmetic floor, aligning with $\mathrm{SO}(3)$ curves.

\subsection*{Training reduces violations; exact zeros require enforcement}

\begin{figure}[!ht]
\centering
\includegraphics[width=0.68\textwidth]{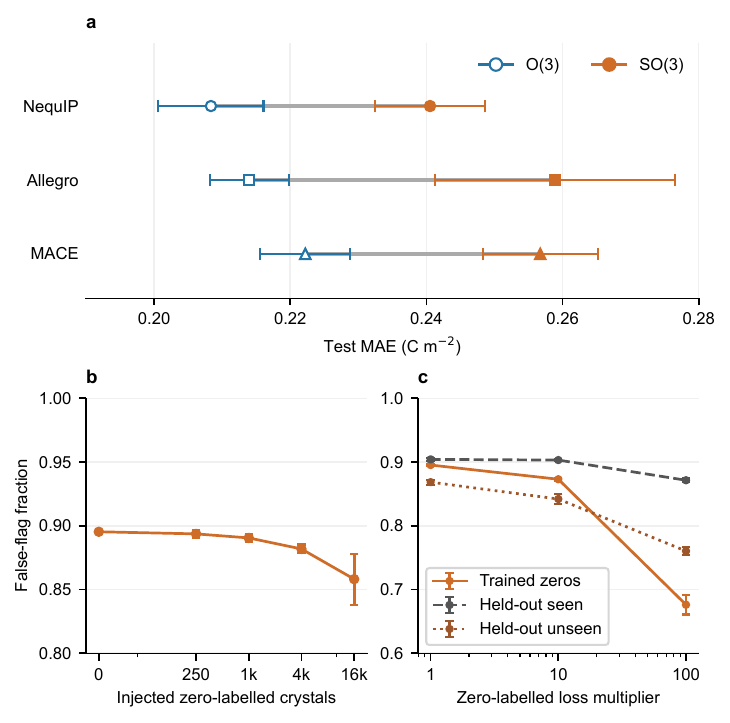}
\caption{\textbf{Accuracy controls and training interventions isolate parity enforcement mechanisms.} \textbf{a}, Piezoelectric regression-test MAE for matched pairs on labeled test data; points represent three-seed means and horizontal bars show standard deviations. Open blue and filled orange markers denote $\mathrm{O}(3)$ and $\mathrm{SO}(3)$ models, respectively, with lines connecting paired arms. Parity enforcement introduces no systematic accuracy penalty. \textbf{b}, Held-out false-flag fractions for NequIP $\mathrm{SO}(3)$ models across zero-labeled data injection sizes (250 to 16,000 crystals). \textbf{c}, False-flag fractions for NequIP $\mathrm{SO}(3)$ models under increasing zero-labeled loss weight multipliers ($1\times$, $10\times$, $100\times$), differentiating trained zero inputs from held-out crystals with space groups seen or unseen during data augmentation. Points and bars denote three-seed means and standard deviations. Data supervision suppresses violation magnitudes but leaves residual false flags above threshold.}
\label{fig:accuracy}
\end{figure}

Predictive accuracy controls across QM9 (parity-even energy, parity-odd dipole)~\cite{ramakrishnan2014qm9} and Materials Project benchmarks (parity-even elasticity, parity-odd piezoelectricity)~\cite{dejong2015piezo} demonstrate that $\mathrm{O}(3)$ parity enforcement introduces no systematic accuracy penalty. On piezoelectric regression tests, $\mathrm{O}(3)$ models achieve equal or lower mean absolute errors (MAE) across all three matched backbones compared to their $\mathrm{SO}(3)$ counterparts (Fig.~\ref{fig:accuracy}a).

Augmenting $\mathrm{SO}(3)$ training sets with 1,000 explicit zero-labeled centrosymmetric crystals reduces violation magnitudes and improves overall regression metrics on held-out space groups. However, soft loss supervision fails to eliminate forbidden predictions: most zero-augmented training inputs continue to exceed the false-flag threshold post-training.

Hyperparameter sweeps over NequIP $\mathrm{SO}(3)$ models varying injected zero counts (250 to 16,000) and loss penalty multipliers (1$\times$ to 100$\times$) confirm the limits of data augmentation (Fig.~\ref{fig:accuracy}b,c). While scaling zero labels or loss weights steadily suppresses violation medians, substantial false-flag fractions persist across all finite sweeps (Supplementary Tables~\ref{stab:augmentation} and~\ref{stab:projection}).

By contrast, explicit output inversion antisymmetrization—a group-averaging transformation $A(x)=\frac{1}{2}[f_\theta(x)-f_\theta(I\!\cdot\!x)]$~\cite{puny2022frame}—guarantees exact physical compliance. For deterministic functions under (A1)--(A2), centrosymmetric input identity enforces $A(x)=0$ identically, eliminating all operating-threshold violations in matched $\mathrm{SO}(3)$ cores. In EqV2, forward-pass and inversion defects prevent complete cancellation, highlighting that output projections rely on exact underlying rotation equivariance.

%% file: sections/discussion.tex
\section*{Discussion}

The parity gap isolates the exact tensor freedom removed by inversion symmetry beyond a crystal's proper rotations, organizing the observed prediction patterns across centrosymmetric space groups. Models restricted to $\mathrm{SO}(3)$ equivariance generate forbidden non-zero outputs specifically within positive-gap classes, while respecting physical zeros only where proper rotations independently enforce them. Under the specified baseline invariances, full $\mathrm{O}(3)$ equivariance mathematically guarantees zero predictions across the entire centrosymmetric population. Evaluating models on symmetry-forced zeros expands the validation of symmetry-aware tensor predictors beyond traditional accuracy benchmarks. Because crystal symmetry dictates the ground-truth zero independent of costly reference-property calculations, physical validity can be evaluated separately from standard regression metrics. Rigorous benchmark protocols must therefore explicitly define input point-group symmetries and output transformation rules, keeping the symmetry-test evaluation set strictly isolated from labeled training and regression data. Data supervision and structural constraints enforce physical laws through fundamentally different mechanisms. While soft loss interventions and zero-label augmentation suppress violation magnitudes, they leave persistent false flags on held-out structures. In contrast, representation-level $\mathrm{O}(3)$ equivariance bakes parity into hidden features to guarantee zero outputs for any parameter values, while post-hoc output antisymmetrization provides an explicit group-averaging mechanism to enforce inversion parity on deterministic forward passes.

\subsection*{Limitations}

Our empirical evaluation is limited to three e3nn-based cores under fixed hyperparameters across three random seeds, where parity reassignments naturally introduce parameter count differences between matched $\mathrm{SO}(3)$ and $\mathrm{O}(3)$ arms. Data augmentation sweeps focus exclusively on NequIP across finite zero-label injections and loss weights, leaving alternative loss formulations and different optimization regimes unexamined. Additionally, numerical coordinate idealization relies on tolerance thresholds, meaning raw-coordinate violations can reflect actual input digital asymmetry, while measured rotation and determinism defects in EquiformerV2 place it outside strict $\mathrm{SO}(3)$ equivariance assumptions. Finally, while polar distortion paths confirm output symmetry transitions, calibrated physical response accuracy along these curves remains outside our experimental scope, and internal feature parity distributions remain unmeasured.

%% file: sections/methods.tex
\section*{Methods}

\subsection*{Targets, coordinates, and reporting threshold}

Piezoelectric target values are defined as the polar tensor $e_{ijk}$, symmetric in its strain indices and measured in units of C\,m$^{-2}$~\cite{dejong2015piezo}. We apply an orthonormal transformation between Cartesian components and the eighteen-dimensional irreducible tensor basis to preserve the Frobenius norm. On centrosymmetric geometries, symmetry mandates a vanishing physical response ($e_{ijk} = 0$)~\cite{klapper2006point}; thus, any non-zero predicted norm represents pure prediction error. To quantify physical violations, we define a \emph{false flag} as a predicted norm exceeding an operating threshold $\tau$. This threshold is calibrated against the empirical training distribution across 25 logarithmically spaced values from $10^{-4}$ to $1$\,C\,m$^{-2}$, serving as an evaluation cutoff rather than a physical measurement boundary.

Crystal point groups and coordinate geometries are identified using spglib. Candidate materials are drawn from the 92 centrosymmetric space groups, filtering for insulators with Materials Project band gaps exceeding 0.1\,eV. Following a deterministic shuffle, structures undergo re-verification at a symmetry tolerance of \texttt{symprec} $= 10^{-3}$\,\AA, rejecting 103 candidates before retaining the first 2,000 verified materials. Idealized coordinates snap atomic positions strictly to their assigned space group, whereas raw variants preserve density-functional-relaxed geometries. Tolerance-based classifications provide symmetry labels for raw coordinate arrays, with exceptional cases documented alongside robustness controls.

\subsection*{Data and splits}

Crystal structures from the Materials Project are processed using pymatgen~\cite{ong2013pymatgen}. The piezoelectric dataset comprises 3,312 structures partitioned into 2,649/331/332 training, validation, and test examples; elasticity contains 13,080 structures split into 10,464/1,308/1,308. Prior to splitting, four piezoelectric outliers (norm $> 50$\,C\,m$^{-2}$) and 147 failed elasticity calculations (components $> 2,000$\,GPa) were removed. Evaluation identifiers share no overlap with piezoelectric training partitions. Within the complete labeled piezoelectric set, 22 structures have exact zero reference tensors (16 in training), though none are centrosymmetric; proper rotations forbid piezoelectricity for 12 of these 16 training zeros under point group 432.

Molecular data from QM9 contains 130,831 structures after excluding 3,054 uncharacterized molecules, partitioned into 110,000/10,000/10,831 examples. Molecular models are trained on a shared 25,000-example subset, evaluating total internal energy at 0\,K ($U_0$, reported directly in eV without atomic-reference subtraction) and dipole vectors. Dipoles are computed via atomic charge-position summation $\sum_i q_i\mathbf{r}_i$~\cite{schutt2021painn} using reference Mulliken charges converted to debye, satisfying the required polar transformation law. All random partitions, candidate shuffles, and model arms share a single fixed random seed (42).

\subsection*{Matched models and readouts}

Matched $\mathrm{O}(3)$ and $\mathrm{SO}(3)$ arms of NequIP, Allegro, and MACE are constructed through a unified irreducible-representation interface. $\mathrm{O}(3)$ models assign natural spherical-harmonic parity $(-1)^l$ to angular degree $l$, whereas $\mathrm{SO}(3)$ variants relabel edge harmonics and hidden features as parity-even. Output representations follow $0e$ for scalar energy, $1o$ for vector dipoles, $2\times0e\oplus2\times2e\oplus1\times4e$ for rank-4 elasticity, and $2\times1o\oplus1\times2o\oplus1\times3o$ for rank-3 piezoelectricity. $\mathrm{SO}(3)$ arms override odd output labels as parity-even, applying reduced tensor products to enforce physical index symmetries~\cite{geiger2022e3nn}.

A degree-1 tensor head replaces learned charge-position assembly to directly yield dipole predictions. Feature readouts extract predictions from the deepest layer containing non-scalar features and aggregate over atoms (or edges for Allegro). The supplementary EquiformerV2 (EqV2) implementation utilizes $\mathrm{SO}(2)$ convolutions~\cite{passaro2023escn}, remapping degree-major coefficients to standard multiplicity-major ordering as an independent $\mathrm{SO}(3)$ baseline. To evaluate intensive target scaling against extensive atomic summation, paired experiments compare primary summed readouts with recalibrated mean-pooled configurations trained across identical index sets.

\subsection*{Optimization and numerical probes}

Primary model architectures are optimized using Adam~\cite{kingma2015adam} with a fixed learning rate of $2\times10^{-3}$, zero weight decay, mean-squared error loss on normalized targets, three interaction layers, a 5.0\,\AA{} radial cutoff, and 64 hidden features. Optimization schedules apply fixed epoch counts using final-epoch weights: molecular targets use $l_{\max}=2$ (100 epochs, batch size 32), elasticity uses $l_{\max}=2$ (120 epochs, batch size 16), and piezoelectricity uses $l_{\max}=3$ (150 epochs, batch size 16). Targets are normalized by training statistics; scalar outputs are centered and scaled, whereas vector and tensor targets undergo feature scaling only to preserve equivariance assumption (A3).

Primary experiments execute 84 runs across seven architectural arms, four target properties, and three random seeds (using float32 model weights and float64 geometric inputs). Prior to training, structural transformation probes verify feature representations under proper and improper operations. Trained models are further probed across non-centrosymmetric structures to measure rotation, reflection, and forward-pass determinism without encountering near-zero division artifacts. EqV2 models evaluate transformation consistency using seeded random-frame draws.

\subsection*{Interventions and mechanism measurements}

Augmentation experiments supplement each $\mathrm{SO}(3)$ training set with 1,000 idealized, zero-labeled centrosymmetric crystals drawn from space groups 2, 12, 14, 15, 62, and 225 (disjoint from primary training and evaluation benchmarks). Held-out evaluation materials are evaluated across space groups present in augmentation ($n=1,232$) versus unseen groups ($n=768$). Additional NequIP $\mathrm{SO}(3)$ sweeps vary zero-injection scale (250, 1,000, 4,000, or 16,000 nested crystals) and zero-label loss penalty multipliers ($1\times$, $10\times$, $100\times$).

Test-time output enforcement applies explicit inversion antisymmetrization $[f(x)-f(I\!\cdot\!x)]/2$ centered at the cell origin, where rigid-translation invariance renders the choice of inversion origin immaterial for centrosymmetric structures. Continuous polar distortion paths simulate physical symmetry breaking in centrosymmetric rutile TiO$_2$ (space group 136) by displacing atoms along a [001] polar mode ($x(\delta)=x(0)+\delta\Delta x$, space group 102) across 33 dimensionless amplitudes $\delta$ verified via spglib at $10^{-8}$\,\AA.

\subsection*{Statistics and Reproducibility}

\subsubsection*{Statistical Analyses}

Accuracy metrics reflect sample means and standard deviations across three independent initialization seeds per arm. Standardized differences measure performance shifts ($\mathrm{SO}(3)$ minus $\mathrm{O}(3)$ MAE) in units of pooled standard deviations. Structure-level, one-sided Wilcoxon signed-rank tests~\cite{wilcoxon1945ranking} compare predicted output norms across paired crystal inputs between architectural arms within each seed and coordinate variant. Confidence intervals for false-flag rates are calculated using 2,000 percentile-bootstrap resamples~\cite{efron1986bootstrap} with structure indices shared across seeds. Seed-level comparisons evaluate test MAEs via paired signed-rank tests.

\subsubsection*{Reproducibility}

Primary training runs consumed 72.439 GPU-hours. Full execution manifests—recording configuration hashes, split definitions, package versions, hardware specifications, and random seeds—are archived in the supplementary materials. Complete mathematical formalizations of inversion identities, invariant-space constraints, parity-gap identities, and output antisymmetrization are implemented and verified in Lean 4~\cite{demoura2021lean4} using pinned mathlib~\cite{mathlib2020} environments deposited in the source repository.

\subsection*{Use of AI tools}

A large language model assisted with manuscript drafting, copy editing, formatting, and code development. The authors are responsible for reviewing the scientific statements, proofs, analyses, code, figures, and reported results in the final submitted version.

%% file: sections/supplementary.tex
\setlength{\tabcolsep}{3pt}
\makeatletter
\setlength{\@fptop}{0pt}
\setlength{\@fpbot}{0pt plus 1fil}
\makeatother
\suppnote{Parity-gap proof and output enforcement}\label{snote:proofs}

\subsection*{Proof of Theorem~\ref{thm:gap}}

Under assumptions (A1)--(A3), consider crystal point group $G$, its proper rotation subgroup $G^+$, and the polar rank-$r$ representation $D$ defined on physical tensor space $V$. All invariant subspaces are evaluated within $V$. For a centrosymmetric structure centered at the origin, atomic coordinates satisfy $-\mathbf{r}_i=\mathbf{r}_{\pi(i)}+\mathbf{t}_i$ for a species-preserving permutation $\pi$ and lattice vectors $\mathbf{t}_i$. Assumptions (A1)--(A2) eliminate these permutations and lattice shifts, establishing the structural identity $f_\theta(I\!\cdot\!x)=f_\theta(x)$ (where shifting the inversion center introduces a rigid translation likewise removed by A2). When outputs satisfy full $\mathrm{O}(3)$ equivariance for odd $r$, $D(I)f_\theta(x)=-f_\theta(x)$ must hold simultaneously, forcing $f_\theta(x)=0$ identically across all parameter configurations.

For any $R \in G^+$, a space-group operation $\{R \mid \mathbf{v}\}$ preserves crystal structure up to permutation and lattice translations. Invariance conditions (A1)--(A2) enforce $f_\theta(Rx)=f_\theta(x)$ even for non-symmorphic fractional shifts $\mathbf{v}$, while $\mathrm{SO}(3)$ equivariance mandates $f_\theta(Rx)=D(R)f_\theta(x)$, restricting predictions to the invariant subspace $f_\theta(x) \in V^{G^+}$. This derivation relies strictly on translation invariance under arbitrary rigid shifts.

This constrained space is fully attainable among unrestricted equivariant mappings. Fixing $T \in V^{G^+}$ and viewing structures modulo input invariances, we define $f(Rx)=D(R)T$ on the $\mathrm{SO}(3)$ orbit of $x$, setting $f=0$ elsewhere. If $R_1x=R_2x$ within this quotient, then $R_2^{-1}R_1 \in G^+$, ensuring consistent assignment. On the orbit, equivariance yields $f(QRx)=D(Q)D(R)T$, while off-orbit evaluations evaluate to zero on both sides, demonstrating $f(x)=T$. Attainability here applies to unrestricted functions; network capacity, continuity, and architectural constraints can further restrict the realizable space in practice.

Because spatial inversion $I$ is central and belongs to $G$, every group element decomposes via coset expansion as $G = G^+ \sqcup I G^+$. A tensor fixed by $G^+$ is therefore invariant under $G$ if and only if inversion fixes it. Because $D(I)=(-1)^r\mathrm{id}$,
\[
 V^G=\begin{cases}V^{G^+},&r\ \text{even},\\\{0\},&r\ \text{odd},\end{cases}
 \qquad
 \gamma_r(G)=\begin{cases}0,&r\ \text{even},\\\dim V^{G^+},&r\ \text{odd}.\end{cases}
\]
Index permutations commute with $Q^{\otimes r}$, preserving the invariance of physical index-symmetry subspaces. Applying this restriction is essential: point group 432 preserves fully antisymmetric rank-3 tensors, but the piezoelectric strain symmetry condition $e_{ijk}=e_{ikj}$ eliminates them. Class calculations evaluate exact algebraic group closures and Reynolds-projector ranks directly on physical tensor spaces. Numerical test suites independently verify all 32 point-group closures, the eleven centrosymmetric classes, and class 432.

\subsection*{Output enforcement}

For any deterministic function, output inversion antisymmetrization $A(x)=\frac{1}{2}[f(x)-f(I\!\cdot\!x)]$ satisfies $A(I\!\cdot\!x)=-A(x)$ via $I^2=\mathrm{id}$. Applying the centrosymmetric input identity under (A1)--(A2) yields $A(x)=0$ identically. Output antisymmetrization thus guarantees odd parity independent of internal feature representations, provided forward passes remain strictly deterministic across evaluation calls.

\suppnote{Model construction, regression controls, and output audit}\label{snote:construction}

\subsection*{Parity construction and gate}

All matched $\mathrm{SO}(3)$ ablations relabel both edge spherical harmonics and hidden feature parities as even (including Allegro's tensor track). All three matched core architectures utilize the \texttt{e3nn} library, enabling direct internal representation comparisons within a single framework.

Prior to training, a reduced feature probe (16 channels, $l_{\max}=2$) evaluated internal equivariant layers under proper and improper transformations against formal transformation laws. $\mathrm{O}(3)$ implementations passed both rotation and reflection checks. $\mathrm{SO}(3)$ models passed rotation probes but failed reflection checks as expected. MACE symmetric contractions preserve single-precision representation under double-precision casts, requiring single-precision diagnostic tolerances.

Opening additional tensor-product pathways expands parameter counts in production $\mathrm{SO}(3)$ arms relative to $\mathrm{O}(3)$ baselines, yielding parameter ratios from 1.002 to 1.398 across targets. For piezoelectricity, parameter multipliers are 1.037 (NequIP), 1.004 (Allegro), and 1.398 (MACE); EquiformerV2 comprises 1,489,635 parameters with no standalone $\mathrm{O}(3)$ arm. While capacity differences can influence regression performance, algebraic zero enforcement holds independently of model parameters.

\subsection*{Complete regression controls}

Molecular energy targets ($U_0$) represent an under-converged control optimized at a fixed learning rate directly on total internal energy labels, exhibiting the largest seed variance against a target standard deviation of ~1,086\,eV. Differences on dipole and elasticity tasks remain small relative to pooled seed variability, whereas piezoelectric metrics favor $\mathrm{O}(3)$ models across all three backbones. The labeled piezoelectric regression test set comprises 332 structures, strictly disjoint from the centrosymmetric evaluation benchmark.

Dipole labels are constructed from Mulliken charges. Across 125,600 retained molecules with reference dipole magnitudes $\ge 0.5$\,D, this charge proxy underestimates DFT reference magnitudes by a median of 17\% (IQR: 5--27\%) and overestimates on 18\% of inputs. Molecular point-group audits verified candidate symmetry operations via atom matching, correcting three nominal $D_2$ assignments that contained only $C_2$ operations. Verified symmetry mandates a zero dipole for seven of the 10,831 test molecules. Complete regression results are summarized in Supplementary Table~\ref{stab:accuracy}.

\begin{table}[htbp]
\centering
\caption{Regression-test MAE across target properties and model backbones. Values report sample mean and standard deviation over three random seeds. Standardized differences ($\Delta/\sigma$) compute $\mathrm{SO}(3)$ minus $\mathrm{O}(3)$ MAE divided by pooled seed standard deviation (positive values favor $\mathrm{O}(3)$).}
\label{stab:accuracy}
\begingroup
\fontsize{8.3bp}{9.2bp}\selectfont
\setlength{\tabcolsep}{2.4pt}
\renewcommand{\arraystretch}{0.96}
\begin{tabularx}{\textwidth}{@{}l>{\centering\arraybackslash}X>{\centering\arraybackslash}X>{\centering\arraybackslash}X>{\centering\arraybackslash}X@{}}
\toprule
Core & Target & O(3) & SO(3) & $\Delta/\sigma$ \\
\midrule
NequIP & $U_0$ (eV) & \num{53.650} $\pm$ \num{22.503} & \num{51.665} $\pm$ \num{9.189} & \num{-0.116} \\
NequIP & Dipole (D) & \num{0.052} $\pm$ \num{0.003} & \num{0.053} $\pm$ \num{0.003} & \num{0.421} \\
NequIP & Elastic (GPa) & \num{24.332} $\pm$ \num{0.267} & \num{24.486} $\pm$ \num{0.140} & \num{0.725} \\
NequIP & Piezo ($\mathrm{C\,m^{-2}}$) & \num{0.208} $\pm$ \num{0.008} & \num{0.241} $\pm$ \num{0.008} & \num{4.088} \\
Allegro & $U_0$ (eV) & \num{31.076} $\pm$ \num{12.894} & \num{26.981} $\pm$ \num{4.877} & \num{-0.420} \\
Allegro & Dipole (D) & \num{0.075} $\pm$ \num{0.002} & \num{0.076} $\pm$ \num{0.002} & \num{0.580} \\
Allegro & Elastic (GPa) & \num{23.724} $\pm$ \num{0.246} & \num{23.888} $\pm$ \num{0.307} & \num{0.588} \\
Allegro & Piezo ($\mathrm{C\,m^{-2}}$) & \num{0.214} $\pm$ \num{0.006} & \num{0.259} $\pm$ \num{0.018} & \num{3.427} \\
MACE & $U_0$ (eV) & \num{16.763} $\pm$ \num{6.289} & \num{26.090} $\pm$ \num{12.491} & \num{0.943} \\
MACE & Dipole (D) & \num{0.048} $\pm$ \num{0.002} & \num{0.050} $\pm$ \num{0.004} & \num{0.465} \\
MACE & Elastic (GPa) & \num{24.919} $\pm$ \num{0.544} & \num{24.939} $\pm$ \num{0.598} & \num{0.035} \\
MACE & Piezo ($\mathrm{C\,m^{-2}}$) & \num{0.222} $\pm$ \num{0.007} & \num{0.257} $\pm$ \num{0.008} & \num{4.554} \\
EqV2 & $U_0$ (eV) & -- & \num{20.752} $\pm$ \num{5.994} & -- \\
EqV2 & Dipole (D) & -- & \num{0.038} $\pm$ \num{0.001} & -- \\
EqV2 & Elastic (GPa) & -- & \num{35.142} $\pm$ \num{3.115} & -- \\
EqV2 & Piezo ($\mathrm{C\,m^{-2}}$) & -- & \num{0.216} $\pm$ \num{0.010} & -- \\
\bottomrule
\end{tabularx}
\endgroup
\par\smallskip{\notesize EqV2 is an unpaired implementation; standard deviations provide descriptive spread.}
\end{table}

\subsection*{Output audit and EqV2 provenance}

Trained architectures were audited across 25 non-centrosymmetric crystals using median relative mirror and rotation errors to avoid zero-denominator artifacts. Forward-pass determinism was measured as the maximum absolute component difference across five seeded evaluations on five centrosymmetric structures. Matched $\mathrm{O}(3)$ cores exhibit negligible mirror errors, whereas $\mathrm{SO}(3)$ models yield order-unity errors. Matched backbones display small non-zero rotation errors and repeat spreads, whereas EquiformerV2 exhibits larger defects across all audit metrics (Supplementary Data 1).

EquiformerV2 evaluations were conducted using pinned upstream model definitions (commit \texttt{8fe8cbaf8f3c27865b6e28c21db7867e75a107f7}) with the graph constructor relocated to match project specifications. Because its forward pass redraws edge reference frames dynamically, evaluations utilize seeded random-frame draws (five-draw means or shared draws for transformation pairs). The full-population false-flag rate exhibits a draw standard deviation of $5.000\times10^{-4}$.

\subsection*{Compute and environment}

Primary grid models were trained on NVIDIA RTX 5090 GPUs (NequIP, Allegro, EqV2) and RTX PRO 6000 GPUs (MACE), consuming 72.439 GPU-hours (excluding diagnostic, sweep, and pooling runs).

Software dependencies include Python 3.12, PyTorch 2.11.0 with CUDA 12.8, \texttt{nequip} 0.18.0, \texttt{nequip-allegro} 0.8.3, \texttt{mace-torch} 0.3.16, \texttt{pymatgen} 2026.5.4, and \texttt{spglib} 2.7.0, with \texttt{e3nn} pinned to 0.4.4 for MACE and 0.6.0 for NequIP/Allegro. All models use float32 weights and float64 geometric inputs. Materials Project datasets were retrieved on 2026-07-04.

\suppnote{Population, thresholds, and readout robustness}\label{snote:robustness}

\subsection*{Measured threshold curves and distributions}

Threshold curves evaluate false-flag fractions across 25 saved evaluation thresholds. In training data, the 5th percentile tensor norm is 0.045\,C\,m$^{-2}$ and the median is 0.509\,C\,m$^{-2}$, with 98.528\% of training examples exceeding the operating threshold ($\tau=0.01$\,C\,m$^{-2}$). Full threshold sweeps for idealized and raw coordinates are shown in Supplementary Fig.~\ref{sfig:thresholds}.

\begin{figure}[!ht]
\centering\includegraphics[width=0.76\textwidth]{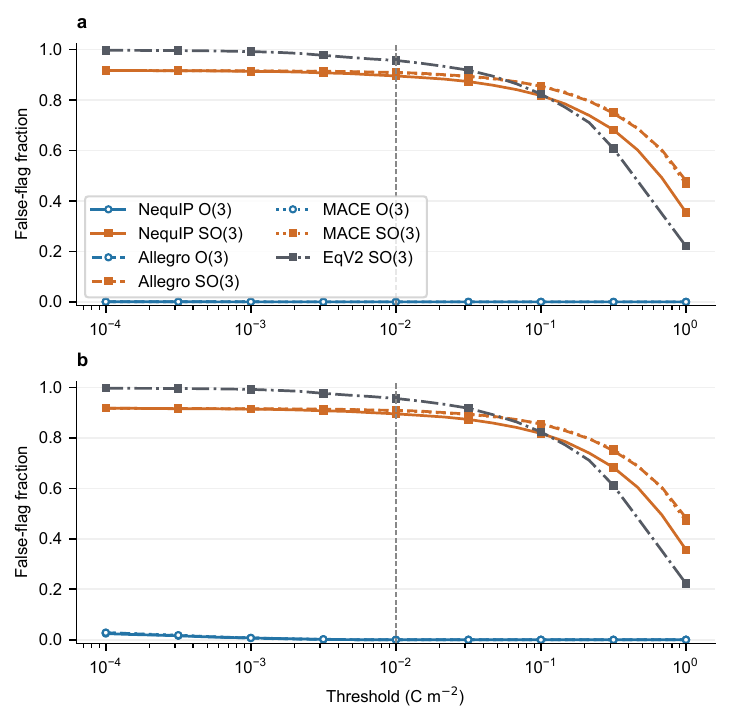}
\caption{False-flag fractions across full threshold sweeps for idealized (\textbf{a}) and raw (\textbf{b}) coordinate variants. Lines connect 25 evaluation thresholds; points indicate seed-mean fractions. Colors distinguish $\mathrm{O}(3)$ from $\mathrm{SO}(3)$ models, and line styles denote core backbones. EquiformerV2 is shown in grey as an unpaired baseline. Dashed vertical lines mark the reporting threshold ($\tau = 0.01$\,C\,m$^{-2}$).}
\label{sfig:thresholds}
\end{figure}

When pooling predictions across structures and seeds, the largest predicted $\mathrm{O}(3)$ norm remains 7.6 times below the operating threshold. MACE models display larger floating-point residual floors than other backbones.

\subsection*{Coordinate tolerance and paired statistics}

All raw structures satisfy centrosymmetry at the initial selection tolerance ($10^{-3}$\,\AA). Under tighter tolerances, 1,956 structures retain centrosymmetric classification at $10^{-4}$\,\AA{} and 1,869 at $10^{-5}$\,\AA{} (compared to 1,999 idealized structures). Idealized crystal \texttt{mp-2923562} fails tight symmetry checks but remains below threshold across all $\mathrm{O}(3)$ runs. Conversely, raw crystal \texttt{mp-1227949} is flagged in three $\mathrm{O}(3)$ runs; its raw coordinates refine to space group 1 at $10^{-4}$\,\AA{} (while idealized coordinates retain space group 47), making non-zero predictions consistent with broken coordinate inversion symmetry.

Evaluating raw coordinates instead of idealized geometries shifts $\mathrm{O}(3)$ false-flag fractions marginally ($+1.667\times10^{-4}$ for NequIP, 0 for Allegro, $+3.333\times10^{-4}$ for MACE). Matched $\mathrm{SO}(3)$ fractions remain unchanged, while EqV2 shifts by $-6.667\times10^{-4}$.

Structure-paired, one-sided Wilcoxon signed-rank tests confirm significantly lower violation norms in $\mathrm{O}(3)$ models across all 18 core/seed/variant combinations ($p < 10^{-300}$ due to library underflow), with $\mathrm{O}(3)$ violation norms lower on 96.2--98.0\% of individual crystals.

\subsection*{Readout scaling and trained pooling comparison}

Supercell scaling tests across nine idealized crystals (space group multiplicities 2, 3, and 8) confirm that non-cubic structures produce extensive outputs under atomic summation and size-invariant outputs under mean pooling, with $\mathrm{O}(3)$ physical zeros persisting under both readouts.

Trained pooling comparisons evaluate dedicated sum-pooled and mean-pooled model pairs using recalibrated target scales and thresholds. All nine matched $\mathrm{O}(3)$ pairs maintain zero false-flag fractions under both readouts. Across twelve $\mathrm{SO}(3)$ configurations, mean pooling shifts false-flag fractions by a mean of $-0.007$ (individual shifts ranging from $-0.038$ to $+0.007$), demonstrating that violation patterns persist under intensive readouts. The paired pooling comparison is summarized in Supplementary Table~\ref{stab:pooling}.

\begin{table}[htbp]
\centering
\caption{Paired trained pooling results evaluated at recalibrated thresholds. Values report three-seed means and standard deviations. Medians denote seed-averaged predicted norms in C\,m$^{-2}$.}
\label{stab:pooling}
\begingroup
\fontsize{8.1bp}{8.9bp}\selectfont
\setlength{\tabcolsep}{2.3pt}
\renewcommand{\arraystretch}{0.94}
\begin{tabularx}{\textwidth}{@{}>{\raggedright\arraybackslash}p{50pt}>{\centering\arraybackslash}X>{\centering\arraybackslash}X>{\centering\arraybackslash}X>{\centering\arraybackslash}X>{\centering\arraybackslash}X@{}}
\toprule
Arm / batch & $F$ mean & $F$ sum & $\Delta F$ & Median mean & Median sum \\
\midrule
Allegro O(3) / 16 & \shortstack{\num{0.000}\\$\pm$ \num{0.000}} & \shortstack{\num{0.000}\\$\pm$ \num{0.000}} & \num{0.000} & \num{2.166e-07} & \num{3.951e-07} \\
MACE O(3) / 16 & \shortstack{\num{0.000}\\$\pm$ \num{0.000}} & \shortstack{\num{0.000}\\$\pm$ \num{0.000}} & \num{0.000} & \num{1.300e-06} & \num{2.465e-06} \\
NequIP O(3) / 16 & \shortstack{\num{0.000}\\$\pm$ \num{0.000}} & \shortstack{\num{0.000}\\$\pm$ \num{0.000}} & \num{0.000} & \num{1.577e-07} & \num{2.430e-07} \\
Allegro SO(3) / 16 & \shortstack{\num{0.902}\\$\pm$ \num{0.002}} & \shortstack{\num{0.898}\\$\pm$ \num{0.002}} & \num{0.004} & \num{0.498} & \num{0.904} \\
EqV2 SO(3) / 16 & \shortstack{\num{0.928}\\$\pm$ \num{0.010}} & \shortstack{\num{0.947}\\$\pm$ \num{0.017}} & \num{-0.019} & \num{0.231} & \num{0.412} \\
MACE SO(3) / 8 & \shortstack{\num{0.897}\\$\pm$ \num{0.001}} & \shortstack{\num{0.898}\\$\pm$ \num{0.001}} & \num{-0.001} & \num{0.406} & \num{0.655} \\
NequIP SO(3) / 8 & \shortstack{\num{0.877}\\$\pm$ \num{0.008}} & \shortstack{\num{0.887}\\$\pm$ \num{0.001}} & \num{-0.011} & \num{0.227} & \num{0.539} \\
\bottomrule
\end{tabularx}
\endgroup
\end{table}

\subsection*{Point-group validation}

The evaluation set contains 166 crystals in class $\mathrm{m}\bar{3}\mathrm{m}$ ($\gamma_3=0$), 18 in class $\mathrm{m}\bar{3}$ ($\gamma_3=1$), and 1,816 non-cubic crystals ($\gamma_3 \in [1, 18]$). Reynolds projections over point group 432 yield rank zero, over 23 yield rank one, and over proper subgroup 422 (rutile) yield rank one. Consequently, proper rotations mandate zero piezoelectricity exclusively for class $\mathrm{m}\bar{3}\mathrm{m}$.

Matched $\mathrm{O}(3)$ false-flag rates are zero across all point-group families and seeds. On zero-gap class $\mathrm{m}\bar{3}\mathrm{m}$, matched $\mathrm{SO}(3)$ models yield zero false flags, whereas EquiformerV2 flags a seed-mean fraction of 0.534 due to implementation-level rotation defects.

\suppnote{Training interventions and output enforcement}\label{snote:repairs}

\subsection*{Augmentation}

Data augmentation incorporates 1,000 zero-labeled centrosymmetric crystals into training, retaining unaugmented target scaling (0.749). Augmentation reduces violation medians and improves regression MAE across all tested $\mathrm{SO}(3)$ cores, but leaves high false-flag fractions on held-out test crystals. Base and augmented arms are compared in Supplementary Table~\ref{stab:augmentation}.

\begin{table}[htbp]
\centering
\captionsetup{font={footnotesize,color=kilslate}}
\caption{Data augmentation effects across $\mathrm{SO}(3)$ and $\mathrm{O}(3)$ model cores on held-out seen ($n=1,232$) and unseen ($n=768$) space groups. $F$ denotes false-flag fraction; medians and test MAE are reported in C\,m$^{-2}$. Two-line cells report three-seed mean and standard deviation.}
\label{stab:augmentation}
\begingroup
\fontsize{7.6bp}{8.3bp}\selectfont
\setlength{\tabcolsep}{2.2pt}
\renewcommand{\arraystretch}{0.94}
\begin{tabularx}{\textwidth}{@{}>{\raggedright\arraybackslash}p{58pt}*{6}{>{\centering\arraybackslash}X}@{}}
\toprule
Arm & $F$ seen & $F$ unseen & \shortstack{Median\\seen} & \shortstack{Median\\unseen} & \shortstack{$F$ trained\\zeros} & Test MAE \\
\midrule
\shortstack[l]{NequIP\\SO(3) base} & \shortstack{\num{0.904}\\$\pm$ \num{0.002}} & \shortstack{\num{0.881}\\$\pm$ \num{0.004}} & \num{0.826} & \num{0.422} & n.m. & \shortstack{\num{0.241}\\$\pm$ \num{0.008}} \\
\shortstack[l]{NequIP\\SO(3) aug.} & \shortstack{\num{0.904}\\$\pm$ \num{0.002}} & \shortstack{\num{0.868}\\$\pm$ \num{0.004}} & \num{0.449} & \num{0.254} & \shortstack{\num{0.895}\\$\pm$ \num{0.002}} & \shortstack{\num{0.228}\\$\pm$ \num{0.002}} \\
\shortstack[l]{NequIP\\O(3) base} & \shortstack{\num{0.000}\\$\pm$ \num{0.000}} & \shortstack{\num{0.000}\\$\pm$ \num{0.000}} & \num{3.319e-07} & \num{2.884e-07} & n.m. & \shortstack{\num{0.208}\\$\pm$ \num{0.008}} \\
\shortstack[l]{Allegro\\SO(3) base} & \shortstack{\num{0.907}\\$\pm$ \num{0.000}} & \shortstack{\num{0.913}\\$\pm$ \num{0.004}} & \num{1.182} & \num{0.586} & n.m. & \shortstack{\num{0.259}\\$\pm$ \num{0.018}} \\
\shortstack[l]{Allegro\\SO(3) aug.} & \shortstack{\num{0.907}\\$\pm$ \num{0.000}} & \shortstack{\num{0.892}\\$\pm$ \num{0.005}} & \num{0.538} & \num{0.274} & \shortstack{\num{0.899}\\$\pm$ \num{0.001}} & \shortstack{\num{0.230}\\$\pm$ \num{0.004}} \\
\shortstack[l]{Allegro\\O(3) base} & \shortstack{\num{0.000}\\$\pm$ \num{0.000}} & \shortstack{\num{0.000}\\$\pm$ \num{0.000}} & \num{4.030e-07} & \num{3.389e-07} & n.m. & \shortstack{\num{0.214}\\$\pm$ \num{0.006}} \\
\shortstack[l]{EqV2\\SO(3) base} & \shortstack{\num{0.951}\\$\pm$ \num{0.005}} & \shortstack{\num{0.967}\\$\pm$ \num{0.007}} & \num{0.543} & \num{0.295} & n.m. & \shortstack{\num{0.216}\\$\pm$ \num{0.010}} \\
\shortstack[l]{EqV2\\SO(3) aug.} & \shortstack{\num{0.930}\\$\pm$ \num{0.005}} & \shortstack{\num{0.935}\\$\pm$ \num{0.011}} & \num{0.149} & \num{0.102} & \shortstack{\num{0.922}\\$\pm$ \num{0.005}} & \shortstack{\num{0.177}\\$\pm$ \num{0.006}} \\
\shortstack[l]{MACE\\SO(3) base} & \shortstack{\num{0.907}\\$\pm$ \num{0.000}} & \shortstack{\num{0.908}\\$\pm$ \num{0.002}} & \num{1.156} & \num{0.508} & n.m. & \shortstack{\num{0.257}\\$\pm$ \num{0.008}} \\
\shortstack[l]{MACE\\SO(3) aug.} & \shortstack{\num{0.907}\\$\pm$ \num{0.001}} & \shortstack{\num{0.892}\\$\pm$ \num{0.002}} & \num{0.534} & \num{0.259} & \shortstack{\num{0.896}\\$\pm$ \num{0.002}} & \shortstack{\num{0.222}\\$\pm$ \num{0.011}} \\
\shortstack[l]{MACE\\O(3) base} & \shortstack{\num{0.000}\\$\pm$ \num{0.000}} & \shortstack{\num{0.000}\\$\pm$ \num{0.000}} & \num{2.974e-06} & \num{2.298e-06} & n.m. & \shortstack{\num{0.222}\\$\pm$ \num{0.007}} \\
\bottomrule
\end{tabularx}
\endgroup
\end{table}

\subsection*{NequIP zero-injection and loss-weight sweeps}

Increasing zero-labeled injection sizes (250 to 16,000 structures) or loss penalty multipliers ($1\times$ to $100\times$) in NequIP $\mathrm{SO}(3)$ models systematically suppresses violation medians and reduces held-out false flags. However, substantial false-flag fractions persist across all finite intervention sweeps.

\subsection*{Output inversion antisymmetrization}

Applying explicit output inversion antisymmetrization reduces false-flag fractions to zero across all matched $\mathrm{SO}(3)$ cores on idealized inputs, enforcing exact zero outputs at numerical precision limits. EquiformerV2 retains residual false flags (0.822) post-projection due to forward-pass stochastic frame sampling. Original and projected false-flag fractions are reported in Supplementary Table~\ref{stab:projection}.

\begin{table}[htbp]
\centering
\captionsetup{font={footnotesize,color=kilslate}}
\caption{Output projection performance across matched models and coordinate variants. Values represent three-seed averages; median norms reported in C\,m$^{-2}$.}
\label{stab:projection}
\begingroup
\fontsize{7.8bp}{8.5bp}\selectfont
\setlength{\tabcolsep}{2.4pt}
\renewcommand{\arraystretch}{0.94}
\begin{tabularx}{\textwidth}{@{}l>{\centering\arraybackslash}X>{\centering\arraybackslash}X>{\centering\arraybackslash}X>{\centering\arraybackslash}X@{}}
\toprule
Arm & Coordinates & $F$ original & $F$ projected & Projected median \\
\midrule
NequIP O(3) & idealized & \num{0.000} & \num{0.000} & \num{2.148e-07} \\
NequIP O(3) & raw & \num{1.667e-04} & \num{1.667e-04} & \num{2.378e-07} \\
NequIP SO(3) & idealized & \num{0.895} & \num{0.000} & \num{3.032e-07} \\
NequIP SO(3) & raw & \num{0.895} & \num{1.667e-04} & \num{3.396e-07} \\
Allegro O(3) & idealized & \num{0.000} & \num{0.000} & \num{3.849e-07} \\
Allegro O(3) & raw & \num{0.000} & \num{0.000} & \num{4.250e-07} \\
Allegro SO(3) & idealized & \num{0.909} & \num{0.000} & \num{6.472e-07} \\
Allegro SO(3) & raw & \num{0.909} & \num{1.667e-04} & \num{7.248e-07} \\
MACE O(3) & idealized & \num{0.000} & \num{0.000} & \num{2.755e-06} \\
MACE O(3) & raw & \num{3.333e-04} & \num{3.333e-04} & \num{3.170e-06} \\
MACE SO(3) & idealized & \num{0.908} & \num{0.000} & \num{2.400e-06} \\
MACE SO(3) & raw & \num{0.908} & \num{1.667e-04} & \num{2.766e-06} \\
EqV2 SO(3) & idealized & \num{0.955} & \num{0.822} & \num{0.028} \\
EqV2 SO(3) & raw & \num{0.955} & \num{0.824} & \num{0.028} \\
\bottomrule
\end{tabularx}
\endgroup
\end{table}